\documentstyle[12pt]{article}

\newcommand{\be}{\begin{equation}}
\newcommand{\ee}{\end{equation}}
\newcommand{\ben}{\begin{enumerate}}
\newcommand{\een}{\end{enumerate}}
\newcommand{\licznik}{\setcounter{equation}{0}}
\newcommand{\nel}{\newline}

\begin{document}
\title{ Application of the discrete quantum--classical system to
the information transfer.}
\author{Janusz Mi\'{s}kiewicz\thanks{e-mail: jamis@ift.uni.wroc.pl} \\
Institute of Theoretical Physics \\
University of Wroclaw \\
pl. Maxa Borna 9 \\
PL -- 50204 Wroclaw Poland }
\maketitle

\begin{abstract}
Discussion about properties of a transmitter is performed. It gives
the possibility to qualify the quality of a transmition. The main idea
of this paper is to investigate the influence of detectors onto the
information transfer. The detector model is described within the open
system theory using the event--enhanced quantum theory. The special
attention is paid to the role of the classical part of the coupled
quantum--classical system. The theory of quantum--classical detectors
is developed and the optimization of the detectors is performed. The
measurement device with the 100\% efficiency is described.
Additionally the detector which efficiency does not depend on the
number of distinguishable states is obtained. A new approach to the
coding problem is proposed. The coding is based on the measurement
events instead of the mathematical features of the quantum theory.
Within this approach the optimum coding is proposed. Finally the whole
system is examined and it is shown that the amount of quantum states
required for a information transfer is about tens.
\end{abstract}

\section{Introduction}
\licznik

The issue of sending information through a quantum system become very 
popular lately \cite{2} --- \cite{15}.

We may set great hopes on this method of the transmition \cite{4} \cite{9} 
\cite{11} e.g. the transmition is said to be proove again eavesdropping 
\cite{9}, there is a well--known "no--cloning theorem" saying that
unknown quantum state cannot be cloned \cite{zurek} \cite{dieks}, as
well as special method wich secure communication between two users
\cite{brassard};
the speed of a transmition could be bigger than in classical
systems \cite{4}; to make channels capacity bigger \cite{11}; to lower
amount of energy required for generating a signal \cite{11}.

This subject has been studied along the following two lines:
{ \alph{enumi}
\begin{enumerate}
\item Investigation of the properties of information channels (i.e. within 
information theory) \cite{10}~\cite{11}~\cite{13}~\cite{14}.
\item Creation and investigation of the quantum systems which may work as a
transmitter (theoretically build systems 
\cite{5}~\cite{6}~\cite{10}~\cite{15} as well as experimental systems
\cite{16} --- \cite{18}
\end{enumerate}}
It should be noticed that most of works are focused on creation and
investigation of pure quantum systems, leaving untouched the question of the 
role of the measurement of the quantum systems. The presence and influence
of the classical system onto the quantum system is skipped, even in the
articles devoted to the role of the measurement. This quastion is very
important, especially when we consider an efficiency of such a detector.
(A very important parameter when we think about receiving
messages.) Such a  approach omit the process of a transmition of the logical
value into the quantum state and the inverse process -- when the quantum 
state is changed into the classical state (the logical symbol).

In this work I pay a special attention to the role and property of the
detectors being used as a part of a transmitter, which generate and detect
signal transmitted through quantum channels.\footnote{ As a detector I
understand here every device which can measure or change the quantum
state.}\footnote{The role of a detector has an important influence into
two important cases: the quantum information transfer and the quantum
data storage. The result obtained in this paper can be transferred to
the case of storage of information and could give an estimation of the
amount of quantum states required by the task.}

The main point of this paper is a qualification of a way of coding, which
fully avail properties of a system as a discrete coding and destignation of 
the form of operators which allow to maximize the efficeinty of the detector 
(e.g. the operators allow to achive the probability of registration of a 
quantum state equal one).

I think that the main novum of the article is an examining of the classical
properties of the quantum -- classical system, which may be used as a
transmitter, basing on the theory of continous measurment. Also I've
proved that the asymtotic efficiency of a detector does not depend on
the number of recognizable states. This may give a great possibility
to encrease the speed of information transfer by the usage of n--state
encoding. In the example of transmitter we see that within the chosen
method of encoding the illegibility of transmition is not very
sensitive to the efficiency of the detector. It is enough to send
about 60 quantum states to achive the confidence level bigger then
55\%. It is done with the detectors with the efficiency 90\% as well
as 45\%.

This paper is organized as follows. The section \ref{math}
mathematically formulates the basic features of the event--enhanced
theory wich are necessary for the later investigation. The section
\ref{coupling} is the main section of this article. The part
\ref{transmition} analyzes the properties of a system from a practical
point of view. The next two parts \ref{parstruc} \ref{codes} define
and describe the transmitter and properties of codes which match the
features of quantum state transmition. The subsections \ref{binary},
\ref{system_y_n}, \ref{multistate} formulate and discuss two possible
ways of encoding classical state. The special care is taken of the
decoding problem. In both cases of encoding the optimum receiver is
designed. Theirs properties and abilities to receive the classical
message are investigated in the subsection \ref{pardetector},
\ref{optimization} and \ref{v1v2} (\ref{pardetector} -- the first
encoding method; \ref{optimization}, \ref{v1v2} -- the second way of
encoding). Finally one of this method is eliminated, because it is too
difficult to avoid errors during applying this method. The explanation
of this elimination is given in the section \ref{time}.
Eventually I designate within the discrete coding the method how to
assign the logical value to measurement events. The section
\ref{multistate} describes the two state detector, which can be used
as a receiver. The designed measuring device has an ability to detect
100\% of sended quantum states. The section \ref{coupling} show also
that the efficiency of the one state and two states detector could be
the same and the appendix \ref{appendix} proof that the efficiency of
the measuring device does not depend on the number of distinguished
states. The section \ref{example} illustrates the developed theory.
There is an example of a system which can be used as a transmitter.
The system consists of two detectors. The first detector (described in
\ref{detector1}) performs the nondemoliation measurement and together
with a source form the sender. The second detector (part
\ref{detector_d2}) illustrates numerically the activity of the system
and compares the work of two kind of detectors (with the efficiency
90\% and 45\% ). Within the model the simplest classical error
correction method is investigated and the number of states required
for a intelligable transmition is considered.
The end of this paper (section \ref{end}) suggest several possibile ways
of alteration of the system allowing to improve some of their features.

\section{Mathematical background}
\label{math}
\licznik

As it was mentioned in the introduction the most suitable way of describing
the influence of a detector onto evolution of a transmitter is the theory of
coupled quantum--classical systems which was proposed and developed by prof. 
Ph. Blanchard and prof. A. Jadczyk \cite{19} \cite{21} \cite{25} \cite{26} 
\cite{27}.

I use this theory on account of:
\begin{itemize}
\item the straightforward description of a classical system (usualy a 
measurment device), which permits an internal evolution of the classical
system;
\item it enable communication of the systems in both directions i.e. :
\begin{itemize}
\item flow of information from quantum to classical system,
\item control of quantum states and processes by classical parameters (as it 
is done in real experiments).
\end{itemize}
\end{itemize}
This presentation is based on the model proposed by prof. Ph. Blanchard and 
prof. A. Jadczyk in \cite{21}.\footnote{ The interested readers may find a
more precise description of the theory in \cite{19} \cite{20} \cite{25}.}
I restrict this presentation to the model where the classical system is a 
discrete one.

The main idea is to describe the evolution of the system as given by completely
positive semigroups. The system is constructed by coupling the classical and 
quantum one. The most natural way of doing this is to use a tensor product 
of classical and quantum spaces. The evolution is governed by completely 
positive semigroups rather then a unitary evolution (by coupling classical
and quantum system I obtain the system with a dissipation). I assume that
the pure states of the quantum system are given by rays in a complex, finite 
or infinite dimentional Hilbert space $H_{q}$. The observable algebra of the 
quantum system is the algebra $A_{q}=L(H_{q})$ of all bounded operators on 
$H_{q}$. The statistical states of the quantum system are given by positive, 
weakly continuous functionals $w$ on $A_{q}$ with $w(I)=1$. Let $S_{q}$ be a
convex set of these states. The elements of $S_{q}$ are positive operators 
on $H_{q}$ of trace $1$. \newline
Let $X_{cl}$ denote the set of pure states of a classical system. I restrict 
the model to the case where $X_{cl}$ is a finite set (with $n+1$ elements).
$S_{cl}$ is a set of statistical states of the classical system, which is 
the space of probability measures on $X_{cl}$. In this case the state 
$P\in S_{cl}$ are $n+1$ tuples $P=(p_{0},\ldots ,p_{n})$, where 
$p_{\alpha} \geq 0, \sum_{\alpha} p_{\alpha} = 1$. The observable algebra 
of the classical system $A_{cl}$ is the abelian algebra of complex function 
on $X_{cl}$, i. e. $A_{cl} \cong C^{n+1}$.\footnote{There is a possibility 
to build the model where the $X_{cl}$ is a infinite set \cite{SQUID}.}
\newline
The total system has as its algebra:
\be
 A_{tot}=A_{q}\otimes A_{cl}=L(H_{q})\otimes C^{n+1}
\ee
It is convinient to realize $A_{tot}$ as a algebra of operators on some 
auxiliary Hilbert space:
\be
 H_{tot} = H_{q} \otimes C^{n+1} 
\ee
The algebra $ A_{tot} $ is then isomorphic to the algebra of block -- 
diagonal matrices $ A=diag(a_{0}, \ldots ,a_{n}) $, whose entries 
$ a_{\alpha} $ are bounded linear operators on $ H_{q} $. \newline
There are four special operations on this algebras.
\begin{itemize}
\item The embeding of the quantum and classical algebras into $ A_{tot} $.
They are respectively:
\be
 i_{q} : a \in L(H_{q}) \rightarrow a \otimes I=diag_{n+1}(a,\ldots,a)
\ee
\be
i_{c} : \lambda = ( \lambda_{0} , \ldots , \lambda_{n} ) \rightarrow 
diag(\lambda_{0} I , \ldots , \lambda_{n} I ) 
\ee
$ \lambda_{\alpha} \in C $ \newline
so the states of $ A_{tot} $ are represented by block -- diagonal matrices
\be
\rho = diag(\rho_{0}, \ldots ,\rho_{n} ) 
\ee
$ \rho_{\alpha} \in L( H_{q} ) $  
and  
$ \sum_{\alpha} Tr ( \rho_{\alpha} ) = 1 $
\newline
For the expectation value of an obserwable $A\in A_{tot}$ in a state $ 
\Omega \in S_{tot} $ I have $ \Omega (A) = \sum_{\alpha} 
Tr( w_{\alpha} a_{\alpha} ) $. I shall identify states $\Omega$ with
operators $ W $ representing them.
\item The projectors, which projects
states of $ A_{tot} $ onto the states of $ A_{q} $ and $ A_{cl} $.
They are defined as follows:
\begin{eqnarray}
& &  \pi_{q} (\rho) = \sum_{\alpha} \rho_{\alpha}
\label{qrzut}   \\
& &  \pi_{c} (\rho) = ( Tr ( \rho_{0} ), \ldots , Tr ( \rho_{n} ) )
\label{crzut} \\
\end{eqnarray}
Thus
\be
 Tr ( \rho \, \, i_{q} ( a ) ) = Tr ( \pi_{q} ( \rho ) a )
\ee
and
\be
 Tr ( \rho \, \, i_{c} ( \lambda ) ) = \sum_{ \alpha } \pi_{ c } (
\rho )_{ \alpha }  \lambda_{ \alpha } 
\ee
\end{itemize} 
If we have states $ P = ( p_{0}, \ldots , p_{n} ) \in S_{ cl } $ and 
$ w \in S_{ q } $ the state of the joint system may be build in the way:
\be
 w \otimes P = diag ( p_{0} w , \ldots , p_{n} w ) 
\ee
This is very useful because it allows us to build an initial state. This
state represents the situation when there are no correlations between the
states. As I said the time evolution was given by completely positive
semigroups.\footnote{A brief discussions why this mathematical structure is
used here can be found in \cite{20} \cite{23} \cite{28} \cite{29}}
\newline
It is a CP (completely positive) semigroup $ \alpha^{t} , t \geq 0
$ of CP maps $ \alpha^{t} $ of the algebra of observables. The time
evolution of states is given by the one parameter semigroup of dual maps 
$ \alpha_{t} : S_{tot} \rightarrow S_{tot} $ with:
\be
 \alpha_{t} ( \rho ) ( A ) = \rho ( \alpha^{ t } ( A ) ) 
\ee
It follows directly from the definition that $ \alpha_{ t } $ maps states
into states, preserving their positivity and normalisation. Owing to the 
theorems by Stinespring and Lindblad \cite{23} \cite{24} any norm continuous 
semigroup of CP maps $ \alpha^{t} $ must be of the form
\be
 \alpha^{t} = \exp( tL ) 
\ee
with
\be
 L(A)=i [ H,A ] + \sum_{i=1}^{N} V_{i} A V_{i}^{\ast} - 
\frac{1}{2} \{ \sum_{i} V_{i} V_{i}^{\ast} , A \} 
\label{a}
\ee
where
\be
 \sum_{i} V_{i} V_{i}^{\ast} \in A_{tot} 
\label{war1}
\ee
and 
\be
 V_{i} A V_{i}^{\ast} \in A_{tot} \: \mbox{whenever} \: A \in A_{tot}
\label{war2}
\ee
$V_{i}$ will be denominated as a coupling operators, because they define
the coupling between classical and quantum system.
$ H $ is an arbitrary Hermitan operator in $ A_{tot} $ : $ H = H^{\ast} \in 
A_{tot} $\footnote{ It is important to observe that the operators $ V_{i} $ 
do not need to belong to the $ A_{tot} $.}.
Let me denote $ \rho(t) = \alpha_{t} ( \rho ) $, so the evolution [\ref{a}]
of observables of the system leads to the Liouville evolution equation for 
states:
\be
 \dot{\rho} (t) = -i [ H, \rho(t) ] + \sum_{i} V_{i}^{\ast} \rho(t) V_{i} -
\frac{1}{2} \{ \sum_{i} V_{i} V_{i}^{\ast} , \rho (t) \}
\label{rown1}
\ee

\section{Optimization of the coupling operators.}
\licznik
\label{coupling}
\subsection{Features of the transmition.}
\label{transmition}
\subsubsection{Speed of the transmition.}

I define this quantity as follows:
\be
\eta = \frac{\mbox{size of the signal}}{\mbox{time of the transmition}}
\ee
{\it Size of the signal} is measured as a number of bits in binary coding.
\newline
This measure is beeing used by engineers when they describe classical 
transmitions (for instance in computer nets).

\subsubsection{Intelligibility of the transmition.}
\label{par4}
This is one of the most important features of the quantum transmition.

The basic features, which influence on the intelligiability are:

\begin{enumerate}
\item resistance to errors, 
\item ability to generate the require signal (for a sender),
\item ability to receive sended signal (for receiver).
\end{enumerate}
This quality may be defined as follows:
\be
\zeta =
\frac{\mbox{input signal} - \mbox{output signal}}{\mbox{size of the signal}}
\ee
Difference between input and output signal is measured as a number of bits 
which are different in input and output information. \newline
Within this part of discussion I want to point out some important
questions, 
which should be answered when the particular system is exemined.
\ben
\item What is the probability of occurance of errors during the transmition?
\item Does this probability depend on the kind of transmitted signal?
\item What is the probability of the proper generation of the quantum state?
\item What is the probability of the proper decoding of the generated quantum
state?
\een

\subsubsection{Confidence level.}

I want to point out here the question about resistance to eavesdropping, 
what is very important from a practical point of view. In which way the
output will change in the presence of the eavestdropping (the increase of 
noise, what kind of noise and so on).

\subsubsection{Resistance to noise and disturbance.}

This can be defined as a ratio of the properly received signals to the 
amount of all signals. Of course the best value is one, because this is the 
case of the ideal receiver.

\subsection{Structure of the transmitter.}
\label{parstruc}
\licznik

The simplest version of the applicable system is:
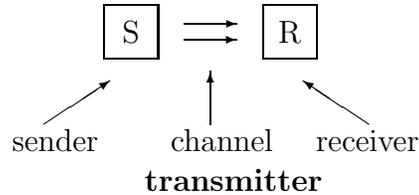
\newline
\begin{picture}(250,100)(0,0)
\put(120,50){\framebox(20,20){S}}
\put(180,50){\framebox(20,20){R}}
\put(150,63){\vector(1,0){20}}
\put(150,57){\vector(1,0){20}}
\put(85,15){sender}
\put(97,25){\vector(3,2){25}}
\put(145,15){channel}
\put(160,25){\vector(0,1){20}}
\put(200,15){receiver}
\put(220,25){\vector(-3,2){25}}
\put(135,0){\bf transmitter}
\end{picture}
\begin{description}
\item{{\bf Sender --}} the quantum--classical system able to generate quantum
state, which may be discerned by receiver. This system 
should generate $n$ distinguishable quantum states. (It makes possibile to 
use $n$ -- state encoding.)
\item{{\bf Receiver --}} the quantum--classical system,
which can distinguish the quantum states generated by sender
end matching the logical value.
\item{{\bf Channel --}} the quantum system transmitting the quantum
signal between sender and receiver\footnote{Within this article I leave the
influence of the channels onto the action of the transmitter, because I want
to concentrate on the property of the sender and receiver. In other words I
assume that the channel is ideal.}.
\end{description}

\subsection{Usage of codes.}
\label{codes}
\licznik

In the following I consider the case of discrete coding. This is the most 
natural way of transmitting information thorough the quantum channels. The 
reason of such a situation is a measurement of a quantum system. As a result
of such a measurement we do not obtain unequivocal result, but
one value is chosen from the all possible results (with some
probability). Consequences of this kind of measurement are very important
for chosing the proper way of encoding. Therefore it is extremally difficult
to construct the system transmitting information in the other way of
encoding -- the continuous encoding (sometimes it is called as a analogue
encoding). This method would require a very sophisticated error correction
system. What more the probability of obtaining the value of the "quantum
variable" with infinite precision is zero. Of course the solution is very
simple -- divide the range of the variable into intervals and assign them
the logical value, but it means to come back to discrete encoding. In
addition, considering the practical requirements of the transmitter it is
obvious that the discrete code will satisfy them.

\subsection{Transmission of the binary signal.}
\label{binary}
\licznik

Accomplishment of the binary code demand to distinguish two different
states of the quantum system, which can be understand as a logical value 
\newline (respectively 0 or 1).

This condition fulfil two kinds of systems:
\begin{enumerate}
\item We make use of the following measurement events:
\begin{itemize}
\item {\bf no registration} of the quantum state -- $0$,
\item {\bf registration} of the quantum state -- $1$.
\end{itemize}
\item The detector descern quantum states and assign the logical value to 
the quantum state.
\end{enumerate}

\subsection{Investigation of the  properties of the system based on
the encoding: registration and no registration.}
\label{system_y_n}
\licznik
\subsubsection{Requirements of the system.}
\label{requirements}

The main problem arise out of the events when the state is properly 
generated but is not registered. It is very difficult to find out that this 
kind of the error has been occured. Moreover the probability of this error 
became greater when the information is longer.

So I have got two following condition:
\begin{itemize}
\item the probability of registration of the quantum state should be $P 
\cong 1$,
\item the work of the sender should be fully controlled.
\end{itemize}
\subsubsection{Building up the receiver.}
\label{buliding}

Firstly I check the conditions for building up the receiver. The main point
of the construction of the receiver is to realize what kind of events can be 
understand by observer. I have chosen two events: registration and no 
registration of the quantum state. \newline
It is possible to use two kinds of the detectors: the classical state shows
the fact of registration of the quantum state or the classical state shows the
number of the registrated quantum states, but the last case can be identified
with a system of $n$ detectors changing their states one by one as the 
registration appears, so the first system will be investigated.
\subsubsection{Description of the receiver.}
\label{pardetector}
The classical state is a probability space. Let $\alpha$ denotes the event 
and $p_{\alpha}$ the probability of the event $\alpha$, then from the 
probability theory:
\be
\sum_{\alpha} p_{\alpha} =1
\label{jedynka}
\ee
In the case of the detector there are two elementary events: registration of
the state and the complementary event (no registration of the quantum 
state).

{\bf Remark} \newline
It should be noticed that the complementary event plays the special role. 
It is natural to make as a initial state
\be
p_{0} ( t=0 ) = 1
\label{pocz}
\ee 
It means that the initial state of the detector is fully known.
It is obvious that in the case of more complicated systems the
complementary event is only one, so should play a special role.

I will use the notation: \newline
$p_{0}$ -- the probability of the complemantary events, \newline
$p_{1}, \ldots ,p_{n}$ -- the probability that $1, \ldots ,n$ event has 
occured. \newline
It follows from the above consideration that the system which
distinguishes $n$ different states should be described within
$n+1$ -- dimentional probability space.

The quantum space is described as a $k$ -- dimentional $( k \geq 1 )$
Hilbert space.

Summing up the assumption I have:
\begin{itemize}
\item $2$ -- dim classical space,
\item $k$ -- dim quantum space,
\item \be \lim_{t \rightarrow \infty} P = 1
      \label{con1}
      \ee
      ($P$-- is the probability of registration of generated quantum state.)

\end{itemize}
{\bf The total space.} \newline
Following the section [\ref{math}]: \newline
$A_{q}$ -- algebra of the quantum system, \newline
$A_{c}$ -- algebra of the classical system. \newline
So the algebra of the total system is: \newline
\be
 A_{tot} = A_{q} \otimes A_{c} = L(H_{q}) \otimes C^{2} 
\ee
The states of the system are block -- diagonal matrixes: 
\be
 \rho \in S_{tot} \, \rho = diag( \rho_{0} , \ldots , \rho_{n} ) 
\label{postro}
\ee
where $ \rho_{0} , \ldots , \rho_{n} \in L(H_{q}) $ \newline
Following the assumption [\ref{pocz}]:
\be
\sum_{i=0}^{n} Tr \rho_{i} = 1
\ee
\be
\forall_{i} \; Tr \rho_{i} \geq 0
\ee
The evolution of the states is described by the equation [\ref{rown1}].
Because of the assumption about the classical space operators $V_{i}$ are:
\be 
V_{i} = \left( \begin{array}{cc}
                a_{i} & b_{i} \\
                c_{i} & d_{i}
               \end{array} \right)
\label{postvi}
\ee  
The operators $V_{i}$ should satisfy the conditions [\ref{war1}],
[\ref{war2}] and using the form of $V_{i}$ [\ref{postvi}] I obtain
\be
V_{i}V_{i}^{\ast} = \left( \begin{array}{cc}
a_{i}a_{i}^{\ast} + b_{i}b_{i}^{\ast} & 
a_{i}c_{i}^{\ast} + b_{i}d_{i}^{\ast} \\
c_{i}a_{i}^{\ast} + d_{i}b_{i}^{\ast} &
c_{i}c_{i}^{\ast} + d_{i}d_{i}^{\ast}
                           \end{array} \right) 
\ee
\be
V_{i}^{\ast} \rho (t) V_{i} = \left( \begin{array}{cc}
a_{i}^{\ast} \rho_{0} (t) a_{i} + c_{i}^{\ast} \rho_{1} (t) c_{i} & 
a_{i}^{\ast} \rho_{0} (t) b_{i} + c_{i}^{\ast} \rho_{1} (t) d_{i} \\
b_{i}^{\ast} \rho_{0} (t) a_{i} + d_{i}^{\ast} \rho_{1} (t) c_{i} &
b_{i}^{\ast} \rho_{0} (t) b_{i} + d_{i}^{\ast} \rho_{1} (t) d_{i}
                           \end{array} \right) 
\ee
and finally:
\be
\left\{ \begin{array}{l}
\sum_{i} a_{i}c_{i}^{\ast} + b_{i}d_{i}^{\ast} = 0 \\
\sum_{i} c_{i}a_{i}^{\ast} + d_{i}b_{i}^{\ast} = 0 \\
\sum_{i} a_{i}^{\ast} \rho_{0} (t) b_{i} + c_{i}^{\ast} \rho_{1} (t) d_{i} = 0 \\
\sum_{i} b_{i}^{\ast} \rho_{0} (t) a_{i} + d_{i}^{\ast} \rho_{1} (t) c_{i} = 0
        \end{array}
\right.
\ee
Reducing the conjugate equation:
\be
\left\{ \begin{array}{l}
\sum_{i} a_{i}c_{i}^{\ast} + b_{i}d_{i}^{\ast} = 0 \\
\sum_{i} a_{i}^{\ast} \rho_{0} (t) b_{i} + c_{i}^{\ast} \rho_{1} (t) d_{i} = 0
        \end{array}
\right.
\label{r1}
\ee
If we consider the system which is investigated, we see that the result of 
the action of this system depends on the initial state and with the assumption
that the operators $V_{i}$ do not depend on the time it is obvious that 
$\rho_{0}$ and $\rho_{1}$ should be eigenvectors of the operators $a_{i},
\, b_{i}, \, c_{i}, \, d_{i}$. \newline
I assume for the simplicity $a_{i}, \, b_{i}, \, c_{i}, \, d_{i}$ are 
mutualy ortogonal (in relation to the parameter $i$). \newline
The equation [\ref{r1}] takes the form:
\be 
\forall_{i} \;
\left\{ \begin{array}{l}
 a_{i}c_{i}^{\ast} + b_{i}d_{i}^{\ast} = 0 \\
 a_{i}^{\ast} \rho_{0} (t) b_{i} + c_{i}^{\ast} \rho_{1} (t) d_{i} = 0
        \end{array}
\right.
\label{r2}
\ee
With the assumption that [\ref{r2}] should be fulfilled $ \forall \rho_{0} 
(t), \, \rho_{1} (t) $ I obtain  the condition:
\be
((a=0 \vee b=0) \wedge (c=0 \vee d=0)) \wedge 
((a=0 \vee c=0) \wedge (b=0 \vee d=0))
\label{r3}
\ee
The condition [\ref{r3}] gives the following possibilities:
\begin{eqnarray}
a=0, \, d=0 & \Rightarrow & V= \left( \begin{array}{cc}
                                   0 & b \\
				   c & 0
				   \end{array}
                           \right)
\label{c1} \\
a=0, \, b=0, \, c=0 &
\Rightarrow  & V= \left( \begin{array}{cc}
                                   0 & 0 \\
				   0 & d
				   \end{array}
                           \right) \\
a=0, \, c=0, \, d=0 &
\Rightarrow & V= \left( \begin{array}{cc}
                                   0 & b \\
				   0 & 0
				   \end{array}
                           \right)
\label{c3} \\
b=0, \, c=0 & \Rightarrow & V= \left( \begin{array}{cc}
                                   a & 0 \\
				   0 & d
				   \end{array}
                           \right) \\
b=0, \, c=0, \, d=0 &
\Rightarrow & V= \left( \begin{array}{cc}
                                   a & 0 \\
				   0 & 0
				   \end{array}
                           \right) \\
a=0, \, b=0, \, d=0  &
\Rightarrow & V= \left( \begin{array}{cc}
                                   0 & 0 \\
				   c & 0
				   \end{array}
                           \right) 
\label{c2} 
\end{eqnarray}
This six instance we may divide into two groups: with the zeros  on the 
diagonal and zeros on the antidiagonal.

The case of the zeros on the antidiagonal is not interesting because all 
equation separates into the independent equations.

In the case of the zeros on the diagonal I will investigate the case
[\ref{c1}] becauce [\ref{c3}] and [\ref{c2}] can be understand as a special 
case of [\ref{c1}]. \newline
The equation [\ref{rown1}] may be rewritten:
\be
\left\{ \begin{array}{l}
\dot{\rho_{0}} (t) = -i [ H, \rho_{0} (t) ] + 
\sum_{i} (c_{i}^{\ast} \rho_{1} (t) c_{i} ) -
\frac{1}{2} \sum_{i} ( \rho_{0} (t) b_{i} b_{i}^{\ast} +  
b_{i} b_{i}^{\ast} \rho_{0} (t)) \\
\dot{\rho_{1}} (t) = -i [ H, \rho_{1} (t) ] + 
\sum_{i} (b_{i}^{\ast} \rho_{0} (t) b_{i} ) -
\frac{1}{2} \sum_{i} ( \rho_{1} (t) c_{i} c_{i}^{\ast} +  
c_{i} c_{i}^{\ast} \rho_{1} (t)) 
\end{array}
\right.
\label{rown2a}
\ee
So the evolution of the classical part of the system is given by
(following \mbox{[\ref{crzut}]}):
\be
\left\{ \begin{array}{l}
Tr \dot{\rho_{0}} (t) = Tr \{ -i [ H, \rho_{0} (t) ] + 
\sum_{i} (c_{i}^{\ast} \rho_{1} (t) c_{i} ) -
\frac{1}{2} \sum_{i} ( \rho_{0} (t) b_{i} b_{i}^{\ast} +  
b_{i} b_{i}^{\ast} \rho_{0} (t)) \} \\
Tr \dot{\rho_{1}} (t) = Tr \{ -i [ H, \rho_{1} (t) ] + 
\sum_{i} (b_{i}^{\ast} \rho_{0} (t) b_{i} ) -
\frac{1}{2} \sum_{i} ( \rho_{1} (t) c_{i} c_{i}^{\ast} +  
c_{i} c_{i}^{\ast} \rho_{1} (t)) \}  
\end{array}
\right.
\label{rown2b}
\ee
Using the properties of the trace:
\be
\left\{ \begin{array}{l}
Tr \dot{\rho_{0}} (t) = - \sum_{i} Tr ( b_{i} b_{i}^{\ast} \rho_{0} (t)
- c_{i} c_{i}^{\ast} \rho_{1} (t) ) \\
Tr \dot{\rho_{1}} (t) = \sum_{i} Tr ( b_{i} b_{i}^{\ast} \rho_{0} (t)
- c_{i} c_{i}^{\ast} \rho_{1} (t) ) 
\end{array}
\right.
\label{rown3}
\ee
We see here a very important feature of the system:
\be 
Tr \dot{\rho_{0}} (t) = - Tr \dot{\rho_{1}} (t)
\ee
Now I consider the range of "$i$". This problem can be solved in two ways: 
to use algebraic methods and check if it is possibile to find such $b$ and
$c$ that: $\sum_{i=1}^{N} b_{i} b_{i}^{\ast} = b b^{\ast}$ and
$\sum_{i=1}^{N} c_{i} c_{i}^{\ast} = c c^{\ast}$. It is a quite simple task. 
If we use the assumption about orthogonality of the operators $b_{i}, \, 
c_{i}$ it is easy to satisfy the above condition by substitution: 
$c= \sum_{i=1}^{N} c_{i} , \, b= \sum_{i=1}^{N} b_{i}$. \newline
The other way of arguing is to see that it is enough to take only one 
operator $V$ to fulfil the condition [\ref{con1}]. So the equation [\ref{rown2a}]
take the form:
\be
\left\{ \begin{array}{l}
\dot{\rho_{0}} (t) = -i [ H, \rho_{0} (t) ] + (c^{\ast} \rho_{1} (t) c ) -
\frac{1}{2} ( \rho_{0} (t) b b^{\ast} + b b^{\ast} \rho_{0} (t)) \\
\dot{\rho_{1}} (t) = -i [ H, \rho_{1} (t) ] + (b^{\ast} \rho_{0} (t) b ) -
\frac{1}{2} ( \rho_{1} (t) c c^{\ast} + c c^{\ast} \rho_{1} (t)) 
\end{array}
\right.
\label{rown4a}
\ee
and the equation [\ref{rown3}]
\be
\left\{ \begin{array}{l}
Tr \dot{\rho_{0}} (t) = - Tr ( c c^{\ast} \rho_{0} (t)
- b b^{\ast}  \rho_{1} (t) ) \\
Tr \dot{\rho_{1}} (t) = - Tr ( c c^{\ast} \rho_{0} (t)
- b b^{\ast}  \rho_{1} (t) ) 
\end{array}
\right.
\label{rown4c}
\ee
I shall use the notation:
\be
\begin{array}{l}
p_{0} (t) =Tr \rho_{0} (t) \\
p_{1} (t) =Tr \rho_{1} (t)
\end{array}
\label{rown4cc}
\ee
so 
\be
\left\{ \begin{array}{l}
\dot{p_{0}} (t) = - Tr ( c c^{\ast} \rho_{0} (t)
- b b^{\ast}  \rho_{1} (t) ) \\
\dot{p_{1}} (t) = - Tr ( c c^{\ast} \rho_{0} (t)
- b b^{\ast}  \rho_{1} (t) ) 
\end{array}
\right.
\label{rown4d}
\ee
{\bf Remark} \newline
The evolution of the system (nontrivial) remains until the balance is
obtained:
\be
Tr \, \, c c^{\ast} \rho_{1} (t) = Tr \, \, b b^{\ast} \rho_{0} (t)
\label{rown4e}
\ee
I assume that the operators $b, \, c$ are the base vector of the quantum 
Hilbert space.\footnote{ It does not affect to the possibility of chosing 
the operators $b, \, c$, because we may change the base vector in order to 
obtain the proper shape of $b, \, c$. \nel
The operators $b, \, c$ may be chosen to fulfil some requirements
e.g. to obtain the maximum of information of quantum states
\cite{levitin}.}
There are two interesting possibility of chosing $b, \, c$:
\ben
\item they differ in coupling constant,
\item they are mutually orthogonal.
\een
Solution of the case [b] require knowlage of the Hamiltonian of the system, 
what more exept the case of the nonlinear Hamiltonian the balance of the 
system can not be achived in a finite time. So it is easier to investigate
the system where the operators $b, \,c$ are of the form [a]. \nel
I shall use the notation:
\be
\begin{array}{l}
b=k_{1} e \\
c=k_{2} e
\end{array}
\label{rown5}
\ee
where $e$ satisfy:
\be
\begin{array}{l} 
Tr e=1 \\
e^{\ast} = e = ee
\end{array}
\ee
The operator $e$ may act in two ways:
\be
\begin{array}{l}
e \rho_{0} (t) = \rho_{0} (t) \\
e \rho_{1} (t) = \rho_{1} (t)
\end{array}
\label{rown8}
\ee
or
\be
\begin{array}{l}
e \rho_{0} (t) \neq \rho_{0} (t) \\
e \rho_{1} (t) \neq \rho_{1} (t)
\end{array}
\label{rown8a}
\ee
Firstly I investigate the case [\ref{rown8}]. \nel
Using the notation [\ref{rown4cc}] the equations [\ref{rown4a}] take the form:
\be
\left\{ \begin{array}{l}
\dot{p_{0}} (t) = - (k_{1}^{2} p_{0} (t) - k_{2}^{2} p_{1} (t)) \\
\dot{p_{1}} (t) = k_{1}^{2} p_{0} (t) - k_{2}^{2} p_{1} (t)  
\end{array}
\right.
\label{rown8b}
\ee
Now I obtain the asymptotic form of $p_{0}, \, p_{1}$. The condition for the 
asymptotic solution $(p_{0} ( \infty ), \, p_{1} ( \infty ))$ is:
\be
\dot{p_{0}} = \dot{p_{1}} = 0
\label{rown8c}
\ee
I use [\ref{rown8b}], [\ref{rown8c}] and the assumption [\ref{jedynka}]
to obtain the solutions:
\be
\left\{ \begin{array}{l}
p_{0} (\infty) = \frac{k_{2}^{2}}{k_{1}^{2} + k_{2}^{2}} \\
p_{1} (\infty) = \frac{k_{1}^{2}}{k_{1}^{2} + k_{2}^{2}} 
       \end{array} 
\right.
\ee
The maximal value of the efficiency of the detector is obtained when 
$ k_{2} = 0$ and $ k_{1} \neq 0 $. In the special case $k_{1}=k_{2} \neq 0 
\, \, p_{1}= \frac{1}{2}$. \newline
The case [\ref{rown8a}] \nel
I use the decomposition:
\be
\begin{array}{c}
\rho_{0} = \rho_{0 \parallel} + \rho_{0 \perp} \\
\rho_{1} = \rho_{1 \parallel} + \rho_{1 \perp} 
\end{array}
\ee
where
\be
\begin{array}{cc}
e \rho_{0 \parallel} = \rho_{0 \parallel} & e \rho_{0 \perp} = 0\\
e \rho_{1 \parallel} = \rho_{1 \parallel} & e \rho_{1 \perp} = 0
\end{array}
\label{rown9}
\ee
Applaying [\ref{rown9}] to the equations [\ref{rown4a}], [\ref{rown4c}]:
\be
\left\{ \begin{array}{l}
\dot{\rho_{0 \parallel}} (t) + \dot{\rho_{0 \perp}} (t) =
 -i [ H, \rho_{0 \parallel} (t) + \rho_{0 \perp} (t) ] +
 k_{2}^{2} e \rho_{1 \parallel} e -
\frac{1}{2} ( \rho_{0 \parallel} e + e \rho_{0 \parallel}) \\
\dot{\rho_{1 \parallel}} (t) + \dot{\rho_{1 \perp}} (t) =
 -i [ H, \rho_{1 \parallel} (t) + \rho_{1 \perp} (t) ] +
 k_{1}^{2} e \rho_{0 \parallel} e -
\frac{1}{2} k_{2}^{2} ( \rho_{1 \parallel} e + e \rho_{1 \parallel}) 
\end{array}
\right.
\label{rown9a}
\ee
I assume that the commutator of the Hamiltonian and $\rho$ does not change 
the decomposition of $\rho$.\footnote{It is possibile to allow the 
Hamiltonian to change the decompocition of $\rho$ but this change should be 
negligable in comparison to the influence of the classical part of the 
system.} \newline
The equation [\ref{rown9a}] may be rewritten:
\be
\left\{ \begin{array}{l}
\dot{\rho}_{0 \perp} (t) =  -i [ H, \rho_{0 \perp} (t) ] \\
\dot{\rho}_{1 \perp} (t) =  -i [ H, \rho_{1 \perp} (t) ] \\
\dot{\rho}_{0 \parallel} (t) =  -i [ H, \rho_{0 \parallel} (t) ] +
 k_{2}^{2} \rho_{1 \parallel} -  k_{1}^{2} \rho_{0 \parallel} \\
\dot{\rho}_{1 \parallel} (t) = -i [ H, \rho_{1 \parallel} (t) ] +
 k_{1}^{2} \rho_{0 \parallel} - k_{2}^{2} \rho_{1 \parallel} \\
\end{array}
\right.
\label{rown9b}
\ee
so
\be
\left\{ \begin{array}{l}
Tr \dot{\rho}_{0 \perp} (t) = 0 \\
Tr \dot{\rho}_{1 \perp} (t) = 0
\end{array}
\right.
\label{rown7}
\ee
Let me take the notation:
\be
\left\{ \begin{array}{l}
b_{0} (t) = \dot{\rho}_{0 \perp} (t) \\
b_{1} (t) = \dot{\rho}_{1 \perp} (t) \\
a_{0} (t) = \dot{\rho}_{0 \parallel} (t) \\
a_{1} (t) = \dot{\rho}_{1 \parallel} (t)
\end{array}
\right.
\label{rown6c}
\ee
The condition [\ref{jedynka}] may be rewritten:
\be
b_{0} (t) + b_{1} (t) + a_{0} (t) + a_{1} (t) = 1
\label{rown6}
\ee
I obtain the condition from [\ref{rown8b}]:
\be
k_{1}^{2} a_{0} (\infty ) = k_{2}^{2} a_{1} (\infty )
\label{rown6a}
\ee
applying [\ref{rown6}] to [\ref{rown6a}]:
\be
a_{1} (\infty ) = \frac{k_{1}^{2}}{k_{1}^{2} + k_{2}^{2}} 
( 1- b_{0} (\infty ) - b_{1} ( \infty ) )
\label{rown6b}
\ee
By rewritting the initial condition [\ref{pocz}] in the notation
[\ref{rown6c}] I get:
\be
\left\{ \begin{array}{l}
a_{0} (0) +b_{0} (0) = 1 \\
a_{1} (0) + b_{1} (0) = 0
\end{array}
\right.
\label{rown6d}
\ee
Applying properties of density matrix: \newline
 $ a_{1} (t) \geq 0, \, b_{1} \geq 0 
\Rightarrow a_{1} (0) = b_{1} (0) =0 $. \newline
I receive with condition [\ref{rown7}]: \newline
$ b_{1} (t) = const, \, b_{0} (t) = const $, \newline
so [\ref{rown6b}] take the form:
\be
a_{1} (\infty ) = \frac{k_{1}^{2}}{k_{1}^{2} + k_{2}^{2}} ( 1- b_{0} )
\ee
We see that the efficiency of the detector depends on the initial state of 
the measured quantum system. This is, somehow, obvious, because if the 
device is not sensitive to some parameter it can not register the system 
without this property.

\subsubsection{Time of detection.}
\label{time}

In the previous paragraph I have investigated the efficiency of the detector 
and the resistance to noise. This section is devoted to the time of 
detection. 

In the case of the decomposition [\ref{rown8}] probabilities are governed by
the equations [\ref{rown8b}]. The solution of this equation has the form:
\be
\left\{
\begin{array}{l}
p_{0} (t) = \frac{k_{1}^{2}}{k_{1}^{2} + k_{2}^{2}} 
e^{-(k_{1}^{2} + k_{2}^{2})t } + \frac{k_{2}^{2}}{k_{1}^{2} + k_{2}^{2}} \\
p_{1} (t) = \frac{k_{2}^{2}}{k_{1}^{2} + k_{2}^{2}} 
(1 - e^{-(k_{1}^{2} + k_{2}^{2})t } )
\end{array}
\right.
\label{rownp2}
\ee
If the $\rho$ is of the form [\ref{rown8a}] the solution is:
\be
\left\{
\begin{array}{l}
p_{0} (t) = 
\frac{a k_{1}^{2}}{k_{1}^{2} + k_{2}^{2}} e^{-(k_{1}^{2} + k_{2}^{2})t } +
\frac{a k_{2}^{2}}{k_{1}^{2} + k_{2}^{2}} + b \\
p_{1} (t) = \frac{a k_{2}^{2}}{k_{1}^{2} + k_{2}^{2}} 
(1 - e^{-(k_{1}^{2} + k_{2}^{2})t } )
\end{array}
\right.
\label{rownp3}
\ee
where $ a= a_{0} (0), \, b=b_{0} (0) $. \newline
We see from the solutions [\ref{rownp2}] [\ref{rownp3}] that the  efficiency 
of the detector depends on the time of coupled evolution of the system. We 
see also that by changing the value of $k_{1}, \, k_{2}$ -- the coupling
constants is possible to change the time required to obtain the particular 
efficiency of the detector.

As I have alredy mentioned in the section [\ref{par4}] the most important 
feature of the system is the intelligibility. In this method of coding the
resistance to errors is not good enough. First of all the ideal efficiency
(100\%) is achived after the infinity time of the measurment and it is
possibilie only when the quantum state is the eigenvalue of the operator 
"$e$". This means that the contribution of a noise and
disturbance during the generation and transmition of the signal is not
allowed. Summing up, because of the difficulties: the finite time of
the measurement, errors
arising during generation and transmition of the signal the efficency is
always lower then 100\%. The result of this is that not every signal
generated by sender is detected by receiver this means that some of the 
signal sended as a logical value "$1$" may be understand as a "$0$". This 
kind of errors is very difficult to eliminate and require sophisticated 
error corection methods, but usage of such methods strongly lower speed of
the transmition. What more the probability of occuring this error become 
bigger when the message is longer. The problem described above is not the 
only difficulty of the system. The very important aspect of this method is
a sequential of the signal i.e. if the "no registration" signal is observed
does it maens that one, two or more zeros have been received? We see that
this system has serious difficulties and solving them may require hard work
to do. In the next section I present the other way of coding, which is not
so sensitive for properties of the detector and give us  a chance to build
up properly working system with high level of confidence.

\subsection{Application of the multistate detector to the
information transfer.}
\label{multistate}
\licznik

The multistate detector, in other words the system distinguishing $n$ 
quantum states, is a more applicable system than the detector described in
the section [\ref{pardetector}], because it is comparatively easy to improve 
its efficiency up to requested level. It can be done just by repeating the 
generation of the signal. What more the number of repetition required
to achive high confidence level does not have to be very big.
It gives hope for high speed of transmition. Of course it will
be a small probability of misunderstanding of the signal because
of a noise but this aspect may be improved by introducing a proper filter.

\subsubsection{Two state detector.}
\label{twodetector}

For the sake of simplicity I will investigate the two--state detector, which
is sufficient for receiving binary code. \newline
The system consist of:
\begin{itemize}
\item the classical space which is the 3 -- dim probability space 
 \newline
 $p(t) = (p_{0}(t),p_{1}(t),p_{2}(t))$ with the initial condition
\be
p_{0}(0)=1, \, p_{1}(0)=0, \, p_{2}(0)=0;
\label{rowna9}
\ee
\item the quantum space, which is a Hilbert space.
\end{itemize} 
Following the section [\ref{math}] the states of the coupled system are of 
the form:
\be
\rho (t) = diag ( \rho_{0}(t),\rho_{1}(t),\rho_{2}(t))
\ee
where the initial condition is:
\be
\rho (0) = diag ( p_{0}(t) \rho_{q}, p_{2}(t) \rho_{q}, p_{2}(t) \rho_{q})
\label{rown9aa}
\ee
$\rho_{q} \in H $ -- the initial state of the quantum system.
Operators in the equation [\ref{rown1}] are of the form:
\be
V_{i} = \left( \begin{array}{ccc}
a^{i}_{11} & a^{i}_{12} & a^{i}_{13} \\
a^{i}_{21} & a^{i}_{22} & a^{i}_{23} \\
a^{i}_{31} & a^{i}_{32} & a^{i}_{33} 
\end{array}
\right)
\label{rown14a}
\ee
The entries of $V_{i}$ belongs to the $L(H)$ (as it was explained in section 
[\ref{math}]). From the condition [\ref{war1}] I obtain the equations:
\be
\left\{
\begin{array}{l}
\sum_{i} a^{i}_{11} a^{i \ast}_{21} + a^{i}_{12} a^{i \ast}_{22} + 
a^{i}_{13} a^{i \ast}_{23} = 0 \\
\sum_{i} a^{i}_{11} a^{i \ast}_{31} + a^{i}_{12} a^{i \ast}_{32} + 
a^{i}_{13} a^{i \ast}_{33} = 0 \\
\sum_{i} a^{i}_{21} a^{i \ast}_{11} + a^{i}_{22} a^{i \ast}_{21} + 
a^{i}_{23} a^{i \ast}_{13} = 0 \\
\sum_{i} a^{i}_{21} a^{i \ast}_{31} + a^{i}_{22} a^{i \ast}_{32} + 
a^{i}_{23} a^{i \ast}_{33} = 0 \\
\sum_{i} a^{i}_{31} a^{i \ast}_{11} + a^{i}_{32} a^{i \ast}_{12} + 
a^{i}_{33} a^{i \ast}_{13} = 0 \\
\sum_{i} a^{i}_{31} a^{i \ast}_{21} + a^{i}_{32} a^{i \ast}_{22} + 
a^{i}_{33} a^{i \ast}_{23} = 0 
\end{array}
\right.
\label{rown10}
\ee
I obtain by reducing the conjugate equations:
\be
\left\{
\begin{array}{l}
\sum_{i} a^{i}_{11} a^{i \ast}_{21} + a^{i}_{12} a^{i \ast}_{22} + 
a^{i}_{13} a^{i \ast}_{23} = 0 \\
\sum_{i} a^{i}_{11} a^{i \ast}_{31} + a^{i}_{12} a^{i \ast}_{32} + 
a^{i}_{13} a^{i \ast}_{33} = 0 \\
\sum_{i} a^{i}_{21} a^{i \ast}_{31} + a^{i}_{22} a^{i \ast}_{32} + 
a^{i}_{23} a^{i \ast}_{33} = 0 
\end{array}
\right.
\label{rown11}
\ee
Using the condition [\ref{war2}] and after the reduction of the conjugate 
equations I have:
\be
\left\{
\begin{array}{l}
\sum_{i} a^{i \ast}_{11} \rho_{0} (t) a^{i}_{12} + 
a^{i \ast}_{21} \rho_{1} (t) a^{i}_{22} + 
a^{i \ast}_{31} \rho_{2} (t) a^{i}_{32} = 0 \\
\sum_{i} a^{i \ast}_{11} \rho_{0} (t) a^{i}_{13} + 
a^{i \ast}_{21} \rho_{1} (t) a^{i}_{23} + 
a^{i \ast}_{31} \rho_{2} (t) a^{i}_{33} = 0 \\
\sum_{i} a^{i \ast}_{12} \rho_{0} (t) a^{i}_{13} + 
a^{i \ast}_{22} \rho_{1} (t) a^{i}_{23} + 
a^{i \ast}_{32} \rho_{2} (t) a^{i}_{33} = 0 
\end{array}
\right.
\label{rown12}
\ee
The operators $V_{i}$ should satisfy the equations [\ref{rown11}] and 
[\ref{rown12}]. Following the same argumentation as in the section 
[\ref{pardetector}] equations [\ref{rown11}], [\ref{rown12}] may be 
rewritten:\be
\forall_{i} \; \left\{
\begin{array}{l}
a^{i}_{11} a^{i \ast}_{21} + a^{i}_{12} a^{i \ast}_{22} + 
a^{i}_{13} a^{i \ast}_{23} = 0 \\
a^{i}_{11} a^{i \ast}_{31} + a^{i}_{12} a^{i \ast}_{32} + 
a^{i}_{13} a^{i \ast}_{33} = 0 \\
a^{i}_{21} a^{i \ast}_{31} + a^{i}_{22} a^{i \ast}_{32} + 
a^{i}_{23} a^{i \ast}_{33} = 0  \\
a^{i \ast}_{11} \rho_{0} (t) a^{i}_{12} + 
a^{i \ast}_{21} \rho_{1} (t) a^{i}_{22} + 
a^{i \ast}_{31} \rho_{2} (t) a^{i}_{32} = 0 \\
a^{i \ast}_{11} \rho_{0} (t) a^{i}_{13} +
a^{i \ast}_{21} \rho_{1} (t) a^{i}_{23} + 
a^{i \ast}_{31} \rho_{2} (t) a^{i}_{33} = 0 \\
a^{i \ast}_{12} \rho_{0} (t) a^{i}_{13} + 
a^{i \ast}_{22} \rho_{1} (t) a^{i}_{23} + 
a^{i \ast}_{32} \rho_{2} (t) a^{i}_{33} = 0 
\end{array}
\right.
\label{rown13}
\ee
The equations [\ref{rown13}] should be satified for every 
$ \rho_{0}(t), \, \rho_{1}(t), \, \rho_{2}(t)$. It could happend when:
\ben
\item all nondiagonal entries are zeros, but this case is not interesting 
for applications;
\item the following condition are satisfied:
\be
\begin{array}{cccccc}
(a_{31} = 0 \vee a_{23} = 0 ) & \wedge &
(a_{21} = 0 \vee a_{23} = 0 ) & \wedge &
(a_{12} = 0 \vee a_{13} = 0 ) & \wedge \\
(a_{12} = 0 \vee a_{32} = 0 ) & \wedge &
(a_{23} = 0 \vee a_{13} = 0 ) & \wedge &
(a_{13} = 0 \vee a_{23} = 0 ) 
\end{array}
\label{rown13a}
\ee
\een
The condition [\ref{rown13a}] gives the eleven possible shapes of the 
operator $V$:
\begin{eqnarray}
a_{31} = a_{23} = a_{12} = 0 & \Rightarrow &
\left(
\begin{array}{ccc} 
0 & 0 & a \\
b & 0 & 0 \\
0 & c & 0
\end{array}
\right)
\label{w1}  \\
a_{32} = a_{21} = a_{13} = 0 & \Rightarrow &
\left(
\begin{array}{ccc} 
0 & a & 0 \\
0 & 0 & b \\
c & 0 & 0
\end{array}
\right)
\label{w2}  \\
a_{32} = a_{21} = a_{12} = 0 & \Rightarrow &
\left(
\begin{array}{ccc} 
0 & 0 & a \\
b & 0 & 0 \\
c & 0 & 0
\end{array}
\right)
\label{w3}  \\
a_{32} = a_{23} = a_{13} = 0 & \Rightarrow &
\left(
\begin{array}{ccc} 
0 & a & 0 \\
b & 0 & 0 \\
c & 0 & 0
\end{array}
\right)
\label{w4}  \\
a_{31} = a_{21} = a_{12} = a_{23} = 0 & \Rightarrow &
\left(
\begin{array}{ccc} 
0 & 0 & a \\
0 & 0 & 0 \\
0 & b & 0
\end{array}
\right)
\label{w5}  \\
a_{31} = a_{21} = a_{12} = a_{13} = 0 & \Rightarrow &
\left(
\begin{array}{ccc} 
0 & 0 & 0 \\
0 & 0 & a \\
0 & b & 0
\end{array}
\right)
\label{w6}  \\
a_{31} = a_{21} = a_{13} = a_{32} = 0 & \Rightarrow &
\left(
\begin{array}{ccc} 
0 & a & 0 \\
0 & 0 & b \\
0 & 0 & 0
\end{array}
\right)
\label{w7}  \\
a_{31} = a_{23} = a_{13} = a_{12} = 0 & \Rightarrow &
\left(
\begin{array}{ccc} 
0 & 0 & 0 \\
a & 0 & 0 \\
b & 0 & 0
\end{array}
\right)
\label{w8}  \\
a_{31} = a_{23} = a_{13} = a_{32} = 0 & \Rightarrow &
\left(
\begin{array}{ccc} 
0 & a & 0 \\
b & 0 & 0 \\
0 & 0 & 0
\end{array}
\right)
\label{w9}  \\
a_{32} = a_{21} = a_{12} = a_{23} = 0 & \Rightarrow &
\left(
\begin{array}{ccc} 
0 & 0 & a \\
0 & 0 & 0 \\
b & 0 & 0
\end{array}
\right)
\label{w10}  \\
a_{32} = a_{21} = a_{12} = a_{23} = 0 & \Rightarrow &
\left(
\begin{array}{ccc} 
0 & 0 & a \\
0 & 0 & 0 \\
b & 0 & 0
\end{array}
\right)
\label{w11}
\end{eqnarray}

\subsubsection{Optimization of the operators $V_{i}$.}
\label{optimization}

Making the allowance for the form of $V_{i}$ [\ref{rown14a}] the equation 
[\ref{rown1}] takes the form:
\be
\left\{
\begin{array}{rcl}
\dot{\rho}_{0} (t) & = & -i [ H, \rho_{0} (t) ] +
\sum_{i} ( a^{i \ast}_{11} \rho_{0} (t) a^{i}_{11} + 
           a^{i \ast}_{21} \rho_{1} (t) a^{i}_{21} + 
           a^{i \ast}_{31} \rho_{2} (t) a^{i}_{31} ) - \\
 & &\frac{1}{2} \{ \sum_{i} ( a^{i}_{11} a^{i \ast}_{11} + 
           a^{i}_{12} a^{i \ast}_{12} + 
           a^{i}_{13} a^{i \ast}_{13} ), \rho_{0} (t) \} \\
\dot{\rho}_{1} (t) & = & -i [ H, \rho_{1} (t) ] +
\sum_{i} ( a^{i \ast}_{12} \rho_{0} (t) a^{i}_{12} + 
           a^{i \ast}_{22} \rho_{1} (t) a^{i}_{22} + 
           a^{i \ast}_{32} \rho_{2} (t) a^{i}_{32} ) - \\
& & \frac{1}{2} \{ \sum_{i} ( a^{i}_{21} a^{i \ast}_{21} + 
           a^{i}_{22} a^{i \ast}_{22} + 
           a^{i}_{23} a^{i \ast}_{23} ), \rho_{1} (t) \} \\
\dot{\rho}_{2} (t) & = & -i [ H, \rho_{2} (t) ] +
\sum_{i} ( a^{i \ast}_{13} \rho_{0} (t) a^{i}_{13} + 
           a^{i \ast}_{23} \rho_{1} (t) a^{i}_{23} + 
           a^{i \ast}_{33} \rho_{2} (t) a^{i}_{33} ) - \\
& & \frac{1}{2} \{ \sum_{i} ( a^{i}_{31} a^{i \ast}_{31} + 
                          a^{i}_{32} a^{i \ast}_{32} + 
                          a^{i}_{33} a^{i \ast}_{33} ), \rho_{2} (t) \} 
\end{array}
\right.
\label{rown14}
\ee
and for the classical subsystem:
\be
\left\{
\begin{array}{l}
\dot{p_{0}} (t) = \sum_{i} Tr (
- ( a^{i}_{12} a^{i \ast}_{12} + a^{i}_{13} a^{i \ast}_{13} ) \rho_{0} (t) + 
a^{i}_{21} a^{i \ast}_{21} \rho_{1} (t) +
a^{i}_{31} a^{i \ast}_{31} \rho_{2} (t) ) \\

\dot{p_{1}} (t) = \sum_{i} Tr (
a^{i}_{12} a^{i \ast}_{12} \rho_{0} (t) - 
( a^{i}_{21} a^{i \ast}_{21} + a^{i}_{23} a^{i \ast}_{23} ) \rho_{1} (t) + 
a^{i}_{32} a^{i \ast}_{32} \rho_{2} (t) ) \\

\dot{p_{2}} (t) = \sum_{i} Tr (
a^{i}_{13} a^{i \ast}_{13} \rho_{0} (t) +
a^{i}_{23} a^{i \ast}_{23} \rho_{1} (t) -
( a^{i}_{31} a^{i \ast}_{31} + a^{i}_{32} a^{i \ast}_{32} ) \rho_{2} (t) )

\end{array}
\right.
\label{rown15}
\ee
Taking under consideration the forms of the operators $V$ [\ref{w1} -- 
\ref{w11}], the equations [\ref{rown15}] take the form: \newline
(\ref{w1})
\be
\left\{
\begin{array}{l}
\dot{p_{0}} (t) = \sum_{i} Tr (
- a^{i}_{13} a^{i \ast}_{13} \rho_{0} (t) + 
a^{i}_{21} a^{i \ast}_{21} \rho_{1} (t)) \\

\dot{p_{1}} (t) = \sum_{i} Tr (
- a^{i}_{21} a^{i \ast}_{21} \rho_{1} (t) + 
a^{i}_{32} a^{i \ast}_{32} \rho_{2} (t))  \\

\dot{p_{2}} (t) = \sum_{i} Tr (
a^{i}_{13} a^{i \ast}_{13} \rho_{0} (t) -
a^{i}_{32} a^{i \ast}_{32}  \rho_{2} (t) )

\end{array}
\right.
\label{w12}
\ee
(\ref{w2})
\be
\left\{
\begin{array}{l}
\dot{p_{0}} (t) = \sum_{i} Tr (
- a^{i}_{12} a^{i \ast}_{12} \rho_{0} (t) + 
a^{i}_{31} a^{i \ast}_{31} \rho_{2} (t))  \\

\dot{p_{1}} (t) = \sum_{i} Tr (
a^{i}_{12} a^{i \ast}_{12} \rho_{0} (t) - 
a^{i}_{23} a^{i \ast}_{23} \rho_{1} (t)) \\

\dot{p_{2}} (t) = \sum_{i} Tr (
a^{i}_{23} a^{i \ast}_{23} \rho_{1} (t) -
a^{i}_{31} a^{i \ast}_{31} \rho_{2} (t))

\end{array}
\right.
\label{w13}
\ee
In the equations [\ref{w12}], [\ref{w13}] operator $V$ gives the 
cascade--like connection of the evolution of the probabilities. \newline
(\ref{w3})
\be
\left\{
\begin{array}{l}
\dot{p_{0}} (t) = \sum_{i} Tr (
- a^{i}_{13} a^{i \ast}_{13} \rho_{0} (t) + 
a^{i}_{21} a^{i \ast}_{21} \rho_{1} (t) +
a^{i}_{31} a^{i \ast}_{31} \rho_{2} (t) ) \\

\dot{p_{1}} (t) = \sum_{i} Tr (- a^{i}_{21} a^{i \ast}_{21}  \rho_{1} (t)) \\

\dot{p_{2}} (t) = \sum_{i} Tr (
a^{i}_{13} a^{i \ast}_{13} \rho_{0} (t) -
a^{i}_{31} a^{i \ast}_{31} \rho_{2} (t) )

\end{array}
\right.
\label{w14}
\ee
(\ref{w4})
\be
\left\{
\begin{array}{l}
\dot{p_{0}} (t) = \sum_{i} Tr (
- a^{i}_{12} a^{i \ast}_{12} \rho_{0} (t) +
a^{i}_{21} a^{i \ast}_{21} \rho_{1} (t)) \\

\dot{p_{1}} (t) = \sum_{i} Tr (
a^{i}_{12} a^{i \ast}_{12} \rho_{0} (t) - 
a^{i}_{21} a^{i \ast}_{21} \rho_{1} (t) ) \\

\dot{p_{2}} (t) = \sum_{i} Tr (- a^{i}_{31} a^{i \ast}_{31} \rho_{2} (t) )

\end{array}
\right.
\label{w15}
\ee
(\ref{w5})
\be
\left\{
\begin{array}{l}
\dot{p_{0}} (t) = \sum_{i} Tr (
-a^{i}_{13} a^{i \ast}_{13} \rho_{0} (t) \\

\dot{p_{1}} (t) = \sum_{i} Tr (a^{i}_{32} a^{i \ast}_{32} \rho_{2} (t) ) \\

\dot{p_{2}} (t) = \sum_{i} Tr (
a^{i}_{13} a^{i \ast}_{13} \rho_{0} (t) -
a^{i}_{32} a^{i \ast}_{32} \rho_{2} (t) )

\end{array}
\right.
\label{w16}
\ee
In the above equations [\ref{w14} -- \ref{w16}] there is a probability which 
evolve independently. \newline
(\ref{w6})
\be
\left\{
\begin{array}{l}
\dot{p_{0}} (t) = 0 \\

\dot{p_{1}} (t) = \sum_{i} Tr (
-a^{i}_{23} a^{i \ast}_{23} \rho_{1} (t) + 
a^{i}_{32} a^{i \ast}_{32} \rho_{2} (t) ) \\

\dot{p_{2}} (t) = \sum_{i} Tr (
a^{i}_{23} a^{i \ast}_{23} \rho_{1} (t) -
a^{i}_{32} a^{i \ast}_{32} \rho_{2} (t) )

\end{array}
\right.
\label{w17}
\ee
In the case of the initial condition [\ref{rown9}] the equations [\ref{w15}] 
gives the constant value of probabilities. \newline
(\ref{w7})
\be
\left\{
\begin{array}{l}
\dot{p_{0}} (t) = \sum_{i} Tr (- a^{i}_{12} a^{i \ast}_{12} \rho_{0} (t) ) \\

\dot{p_{1}} (t) = \sum_{i} Tr (
a^{i}_{12} a^{i \ast}_{12} \rho_{0} (t) - 
a^{i}_{23} a^{i \ast}_{23} \rho_{1} (t) ) \\

\dot{p_{2}} (t) = \sum_{i} Tr (a^{i}_{23} a^{i \ast}_{23} \rho_{1} (t) ) 

\end{array}
\right.
\label{w18}
\ee
(\ref{w8})
\be
\left\{
\begin{array}{l}
\dot{p_{0}} (t) = \sum_{i} Tr (a^{i}_{21} a^{i \ast}_{21} \rho_{1} (t) ) \\

\dot{p_{1}} (t) = \sum_{i} Tr (
- a^{i}_{21} a^{i \ast}_{21} \rho_{1} (t) + 
a^{i}_{32} a^{i \ast}_{32} \rho_{2} (t) ) \\

\dot{p_{2}} (t) = \sum_{i} Tr (a^{i}_{32} a^{i \ast}_{32} \rho_{2} (t) )

\end{array}
\right.
\label{w19}
\ee
(\ref{w11})
\be
\left\{
\begin{array}{l}
\dot{p_{0}} (t) = \sum_{i} Tr (a^{i}_{31} a^{i \ast}_{31} \rho_{2} (t) ) \\

\dot{p_{1}} (t) = \sum_{i} Tr (a^{i}_{23} a^{i \ast}_{23} \rho_{1} (t) ) \\

\dot{p_{2}} (t) = \sum_{i} Tr (
a^{i}_{23} a^{i \ast}_{23} \rho_{1} (t) -
a^{i}_{31} a^{i \ast}_{31} \rho_{2} (t) )

\end{array}
\right.
\label{w20}
\ee
The equations [\ref{w18} -- \ref{w20}] have the property that the only two 
entries of the state $\rho$ are included in the equations. \newline
(\ref{w9})
\be
\left\{
\begin{array}{l}
\dot{p_{0}} (t) = \sum_{i} Tr (
- a^{i}_{12} a^{i \ast}_{12} \rho_{0} (t) + 
a^{i}_{21} a^{i \ast}_{21} \rho_{1} (t) ) \\

\dot{p_{1}} (t) = \sum_{i} Tr (
a^{i}_{12} a^{i \ast}_{12} \rho_{0} (t) - 
a^{i}_{21} a^{i \ast}_{21} \rho_{1} (t) ) \\

\dot{p_{2}} (t) = 0

\end{array}
\right.
\label{w21}
\ee
(\ref{w10})
\be
\left\{
\begin{array}{l}
\dot{p_{0}} (t) = \sum_{i} Tr (
a^{i}_{13} a^{i \ast}_{13} \rho_{0} (t) + 
a^{i}_{31} a^{i \ast}_{31} \rho_{2} (t) ) \\

\dot{p_{1}} (t) = 0 \\

\dot{p_{2}} (t) = \sum_{i} Tr (
a^{i}_{13} a^{i \ast}_{13} \rho_{0} (t) -
a^{i}_{31} a^{i \ast}_{31} \rho_{2} (t) )

\end{array}
\right.
\label{w22}
\ee
The last two equations [\ref{w21}], [\ref{w22}] are the most interesting, 
becauce of their simplicity. Theirs action concentrate on the evolution of one
of the probabilities.

\subsubsection{Implementation of the operators $V_{1}$, $V_{2}$ .}
\label{v1v2}

I have not unswered to the question about the range of the parameter $i$. I 
am not going to give a full algebraic analysis, but I show that when $i
\in \{ 1,2 \} $ and $V_{1}$, $V_{2}$ are of the form [\ref{w9}, \ref{w10}]
it is possible to cover the whole range of the efficiency of the detector.

Designating the shape of operators 
$ a_{12} , \, a_{21}, \, a_{13}, \, a_{31} $ 
I take under the consideration the following aspects:
\ben
\item The pairs $ (a_{12} , \, a_{21}) $ and $ (a_{13}, \, a_{31}) $ should 
significantly differ, otherwise $ p_{1} $ and $ p_{2} $ give the similar 
result and distinguishing quantum states could become impossible. This can
make the transmition unreadable. \newline
The best situation is when they are mutually orthogonal.
\item Make the efficiency of the detector close to one as much as
possible.
\een
We should perceive, that the operators [\ref{w9}], [\ref{w10}] 
give the same evolution as a detector considered in the section
[\ref{pardetector}], so I apply the result of this section and investigate 
their joint--action. \newline 
I use the notation:
\be
\begin{array}{cc}
a_{12} = k_{1} e_{2} & a_{21} = k_{2} e_{2} \\
a_{13} = n_{1} e_{3} & a_{31} = n_{2} e_{3}
\end{array}
\label{rown16} 
\ee
where: 
$$
\begin{array}{l}
k_{1} , \, k_{2} , \, n_{1} , \, n_{2} \in \Re \\
( e_{2} , e_{3} ) = 0 \\
 e^{\ast}_{2} = e_{2} = e^{2}_{2} \\
 e^{\ast}_{3} = e_{3} = e^{3}_{2} \\
\end{array} 
$$ 
So the equation [\ref{rown15}] may be rewritten:
\be
\left\{
\begin{array}{l}
\dot{p_{0}} (t) = - Tr ( (k_{1}^{2} e_{2} + n_{1}^{2} e_{3}) \rho_{0} (t) -
k_{2}^{2} e_{2} \rho_{1} (t) - n_{2}^{2} e_{3} \rho_{2} (t) ) \\
\dot{p_{1}} (t) = 
Tr ( k_{1}^{2} e_{2} \rho_{0} (t) -k_{2}^{2} e_{2} \rho_{1} (t) ) \\
\dot{p_{2}} (t) = 
Tr ( n_{1}^{2} e_{3} \rho_{0} (t) - n_{2}^{2} e_{3} \rho_{2} (t) ) \\
\end{array}
\right.
\label{rown17}
\ee
Using the notation [\ref{rown9aa}] I consider the following initial states:
\ben
\item  $ e_{2} \rho_{q} = \rho_{q} $ \newline
       $ e_{3} \rho_{q} = 0 $
\item  $ e_{2} \rho_{q} = 0 $ \newline
       $ e_{3} \rho_{q} = \rho_{q} $ 
\een
\ben
\setcounter{enumi}{2}
\item  $ \rho_{q} = a e_{2} + b e_{3} $ this case could be modificated 
in the way: $ \rho_{q} = \sum_{i} a_{i} e_{i} $  then \newline
\,  $ e_{2} \rho_{q} = a_{2} e_{2} $ \newline
\,  $ e_{3} \rho_{q} = a_{3} e_{3} $.
\een
The case a),b) can be considered jointly, because by changing the notation 
we can transform one case into another obtaning the same equations. What 
more all this cases are included in the c). (We can obtain them by proper 
choice of coefficients $a_{2}, \, a_{3}$.) \newline
From the equation [\ref{rown14}] I obtain:
\be
\left\{
\begin{array}{l}
\dot{\rho}_{0} (t) = - i [H, \rho_{0} (t) ] +
k_{2}^{2} e_{2} \rho_{1} (t) e_{2} + n_{2}^{2} e_{3} \rho_{2} (t) e_{3} -
\frac{1}{2} \{ k_{2}^{2} e_{2} + n_{1}^{2} e_{3} , \rho_{0} (t) \} \\

\dot{\rho}_{1} (t) = - i [H, \rho_{1} (t) ] +
k_{1}^{2} e_{2} \rho_{0} (t) e_{2} -
\frac{1}{2} \{ k_{2}^{2} e_{2} , \rho_{1} (t) \} \\

\dot{\rho}_{2} (t) = - i [H, \rho_{2} (t) ] +
n_{1}^{2} e_{3} \rho_{0} (t) e_{3} -
\frac{1}{2} \{ n_{2}^{2} e_{3} , \rho_{2} (t) \} \\
\end{array}
\right.
\label{rown20}
\ee
Using the decompisition c) and assuming, that only $ a_{2}$, $a_{3}$ 
depend on time and
\be
\begin{array}{l}
\forall_{i \neq 2} \; ( [ H, e_{2} ], e_{i} ) = 0 \, \, and \\
\forall_{i \neq 3} \; ( [ H, e_{3} ], e_{i} ) = 0
\end{array}
\ee
I get
\be
\left\{
\begin{array}{l}
\dot{a_{0}} (t) e_{2} = -i[H, a_{0} (t) e_{2} ] +
( k_{2}^{2} a_{1} (t) - k_{1}^{2} a_{0} (t)) e_{2} \\

\dot{b_{0}} (t) e_{3} = -i[H, b_{0} (t) e_{3} ] +
( n_{2}^{2} b_{2} (t) - n_{1}^{2} b_{0} (t)) e_{3} \\

\dot{a_{1}} (t) e_{2} = -i[H, a_{1} (t) e_{2} ] +
( k_{1}^{2} a_{0} (t) - k_{2}^{2} a_{1} (t)) e_{2} \\

\dot{a_{2}} (t) e_{2} = -i[H, a_{2} (t) e_{2} ] \\

\dot{b_{2}} (t) e_{3} = -i[H, b_{2} (t) e_{3} ] -
( n_{1}^{2} b_{0} (t) - n_{2}^{2} b_{2} (t)) e_{3} \\

\dot{b_{1}} (t) e_{3} = -i[H, b_{1} (t) e_{3} ] \\ 
\end{array}
\right.
\ee
After the trace operation:
\be
\left\{
\begin{array}{l}
\dot{a_{0}} (t) = k_{2}^{2} a_{1} (t) - k_{1}^{2} a_{0} (t) \\

\dot{b_{0}} (t) = n_{2}^{2} b_{2} (t) - n_{1}^{2} b_{0} (t) \\

\dot{a_{1}} (t) = k_{1}^{2} a_{0} (t) - k_{2}^{2} a_{1} (t) \\

\dot{b_{1}} (t) = 0 \\

\dot{a_{2}} (t) = 0 \\

\dot{b_{2}} (t) = n_{1}^{2} b_{0} (t) - n_{2}^{2} b_{2} (t) 
\end{array}
\right.
\label{rown21}
\ee
from [\ref{rown21}] we see:
\be
\begin{array}{l}
a_{2} = const \\
b_{1} = const
\end{array}
\ee
The equation [\ref{rown21}] may be divided into two sets:
\be
\left\{
\begin{array}{l}
\dot{a_{0}} (t) = k_{2}^{2} a_{1} (t) - k_{1}^{2} a_{0} (t) \\
\dot{a_{1}} (t) = k_{1}^{2} a_{0} (t) - k_{2}^{2} a_{1} (t) 
\end{array}
\right.
\label{rown22a}
\ee
and
\be
\left\{
\begin{array}{l}
\dot{b_{0}} (t) = n_{2}^{2} b_{2} (t) - n_{1}^{2} b_{0} (t) \\
\dot{b_{2}} (t) = n_{1}^{2} b_{0} (t) - n_{2}^{2} b_{2} (t) 
\end{array}
\right.
\label{rown22b}
\ee
The solution of the equations [\ref{rown22a}] and [\ref{rown22b}] is:
\be
\left\{
\begin{array}{l}
a_{0} (t) = - A e^{-(k_{1}^{2} + k_{2}^{2}) t} + 
\frac{k_{2}^{2}}{k_{1}^{2}} C \\

a_{1} (t) = A e^{-(k_{1}^{2} + k_{2}^{2}) t} + C \\

b_{0} (t) = B e^{-(n_{1}^{2} + n_{2}^{2}) t} + 
\frac{n_{2}^{2}}{n_{1}^{2}} D \\

b_{2} (t) = - B e^{-(n_{1}^{2} + n_{2}^{2}) t} + D 

\end{array}
\right.
\label{rown23}
\ee
The conditions for the coefficients arrise from [\ref{rown23}],
[\ref{pocz}] and they are:
\be
\left\{
\begin{array}{l}
\frac{n_{2}^{2}}{n_{1}^{2}} D + \frac{k_{2}^{2}}{k_{1}^{2}} C = A+B+1 \\
A+C = 0 \\
B+D = 0
\end{array}
\right.
\label{rown24}
\ee
The exact solution can be found when the initial state is known. I consider 
some important cases:
\ben
\item $D = 0 \Rightarrow \begin{array}{l}
p_{1} (t \rightarrow \infty ) =\frac{k_{1}^{2}}{k_{1}^{2} + k_{2}^{2}} \\
p_{2} (t \rightarrow \infty ) = 0
\end{array} $
The maximal value is obtained for $ k_{1} \neq 0, \, \, k_{2} = 0 $ (like in 
the section [\ref{pardetector}]).
\item $ C = 0 \Rightarrow \begin{array}{l}
p_{1} (t \rightarrow \infty ) = 0 \\
p_{2} (t \rightarrow \infty ) =\frac{k_{1}^{2}}{k_{1}^{2} + k_{2}^{2}}
\end{array} $ \newline
As above the maximum of the efficiency of the detector is obtained for:
\newline
$  n_{1} \neq 0, \, \, n_{2} = 0 $ \newline
It is interesting to see what has happend when non of the components of 
$\rho$ is distinguished:
$$
\begin{array}{l}
C=D \Rightarrow \\
p_{1} (t \rightarrow \infty ) = p_{2} (t \rightarrow \infty ) =
\frac{k_{1}^{2} n_{1}^{2}}{n_{1}^{2}( k_{1}^{2} + k_{2}^{2} ) 
+ k_{1}^{2} ( n_{1}^{2} + n_{2}^{2} ) }
\end{array}
$$
\een
So the maximum is obtained when: \newline
$  k_{1} \neq 0, \; n_{1} \neq 0, \; k_{2} = n_{2} = 0 $
and $ p_{1} (t \rightarrow \infty ) = p_{2} (t \rightarrow \infty ) =
\frac{1}{2} $
\newline
The maximum of the efficiency of the detector for any signal is obtained 
when: \newline
$p_{1} (t \rightarrow \infty ) + p_{2} (t \rightarrow \infty ) = 1 $
\newline
so $ k_{2} = n_{2} = 0 $.

\subsubsection{Conclusion.}
\label{conclusion}

The maximal value of the efficiency depends on the initial state of the 
quantum system (as in the section [\ref{pardetector}]), but it is also
possible to control this value by the choice of the coefficients: $ n_{1},
\; n_{2}, \; k_{1}, \; k_{2} $. The system may be simplified by the 
assumption: $ n_{1} = k_{1}, \; n_{2} = k_{2} $. The efficiency of the 
system may still achive the value one. This choice has one more advantage,
it guarantee the same efficiency for both kind of signal.

\section{Example of the transmitter.}
\label{example}
\licznik

I investigate the simplest transmitter, which use the quantum state as a 
carrier. \newline
I use the following system:
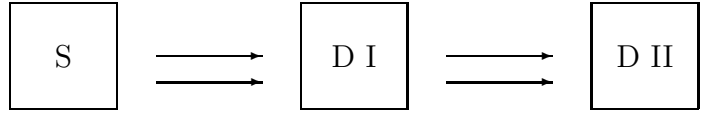
\newline
\begin{picture}(350,60)(0,0)
\put(70,10){\framebox(40,40){S}}
\put(180,10){\framebox(40,40){D I}}
\put(290,10){\framebox(40,40){D II}}
\put(125,20){\vector(1,0){40}}
\put(125,30){\vector(1,0){40}}
\put(235,20){\vector(1,0){40}}
\put(235,30){\vector(1,0){40}}
\end{picture}
\newline
\begin{description}
\item{{\bf S -- }} source of the quantum states;
\item{{\bf D I -- }} the first detector (working also as a filter of a signal);
\item{{\bf D II -- }} the second detector ( it plays the role of the receiver).
\end{description}
In the notation of the section [\ref{parstruc}] the source and detector D I 
together form the sender. \newline
I assume that the source generate the quantum state which has the required 
properties (some level of the noise is allowed). \newline
The detectors D I, D II are the open systems so theirs evolutions is
described by equation [\ref{rown1}]. \newline
I investigate the following initial quantum state:
\ben
\item $ \rho (0) = \rho e$
\item $ \rho (0) = \sum_{i} \rho_{i} e_{i}$
\item $ \rho (0) = \sum_{i} \rho_{i} e_{i} + 
\sum_{i \neq j} \rho_{ij} e_{ij} $
\een
\begin{description}
\item{$e_{i}$ -- } a normalized projector $e_{i} \in L(H)$ (it describes the
diagonal elements);
\item{$e_{ij}$ -- } a projector which describes the non diagonal elements.
\end{description}

\subsection{Properties of the detector D I.}
\label{detector1}

This detector apart from a registration of the states should pass the states 
in unchanged states (unchanged as much as possible). I assume the
following form of the coupling operator $V$:
\be
V = \sqrt{k} \left( \begin{array}{cc}
 0 & e_{1} \\
e_{1} & 0 
\end{array}
\right)
\ee
$\sqrt{k}$ -- coupling constant, \newline
$e_{1}$ -- normalized projector. \newline
For the sake of simplicity I assume $ [H,\rho (t)] = 0$. \newline
This assumption should be understood that the evolution generated by 
Hamiltonian operator must be negligable in comparison to the evolution
governed by the coupling. This is quite easy to achive,
the coupling constant may be increased due to shorter the
influence of the Hamiltonian evolution. \nel
The probability space is a 2 -- dim one,
so the equation [\ref{rown1}] has the solution (following \cite{21}~):
\be
\left\{ \begin{array}{lcl}
\rho_{0} (t) & = & p_{0} ( \frac{1}{2} e \rho_{q} e + f \rho_{q} f ) +
                 \frac{1}{2} p_{1} e \rho_{q} e + \\
& & e^{- \frac{1}{2} k t} p_{0} ( e \rho_{q} f + f \rho_{q} e ) + \\
& & e^{- k t} p_{0} ( e \rho_{q} e + e \rho_{q} e ) + \\
& & \frac{1}{2} e^{-2 k t} ( p_{0} e \rho_{q} e + p_{1} e \rho_{q} e ) \\
\rho_{1} (t) & = & \frac{1}{2} p_{0} e \rho_{q} e + 
   p_{1} ( \rho_{q} + \frac{3}{2} e \rho_{q} e - \{ e, \rho_{q} \} ) + \\
& & e^{- \frac{1}{2} k t} p_{1} ( \{ e, \rho_{q} \} - 2 e \rho_{q} e ) + \\
& & \frac{1}{2} e^{-2kt} (p_{1} e \rho_{q} e - p_{0} e \rho_{q} e)
\end{array}
\right.
\label{rown26}
\ee
where: \newline
$ f = 1 - e $ \newline
$ \forall_{i} \: \rho_{i} (0) = p_{i} \rho_{q} $ \newline
Applying the projector [\ref{qrzut}] to the solution
[\ref{rown26}] I obtain:
\be
\rho_{q} (t) = \rho_{q} + 2 e \rho_{q} e + \{ e, \rho_{q} \} +
e^{ - \frac{1}{2} kt } [ \{ e, \rho_{q} \} - 2e \rho_{q} e ]
\label{rown27}
\ee
In the cases of the initial condition a) -- c) the solution [\ref{rown27}] 
gives
\ben
\item $ \rho_{q} (t) = \rho_{q} $
\item $ \rho_{q} (t) = \rho_{q} $
\item $ \rho_{q} (t) = \sum_{i} \rho_{i} e_{i} + 
\sum_{ij} [ 1+( \delta_{i1} + \delta_{j1} ) ( e^{ - \frac{1}{2} kt } -1)] 
\rho_{ij} e_{ij} $
\een
Applying the projector [\ref{crzut}] onto the classical
subsystem I obtain:
\be
\left\{ \begin{array}{l}
p_{0} (t) = \frac{1}{2} ( p_{1} - p_{0} ) q_{1} + p_{0} +
\frac{1}{2} ( p_{0} - p_{1} ) q_{1} e^{-2kt} \\
p_{1} (t) = \frac{1}{2} ( p_{0} - p_{1} ) q_{1} + p_{1} +
\frac{1}{2} ( p_{1} - p_{0} ) q_{1} e^{-2kt} 
\end{array}
\right.
\label{rown28}
\ee
where \newline
$q_{1} = Tr \; e_{1} \rho_{q} $ \newline
For the initial state [\ref{pocz}], with the assumption [\ref{jedynka}] I 
obtain for the cases a) -- c):
$$
\begin{array}{ll}
\makebox[1cm]{a)} & \left\{ \begin{array}{l}
p_{0} (t) = \frac{1}{2} ( 1 - e^{-2kt} ) \\
p_{1} (t) = \frac{1}{2} ( 1 + e^{-2kt} ) 
\end{array}
\right. \\
\makebox[1cm]{b)} & \left\{ \begin{array}{l}
p_{0} (t) = 1+ \frac{1}{2} ( e^{-2kt} - 1 ) \rho_{1} \\
p_{1} (t) = \frac{1}{2} ( 1 - e^{-2kt} ) \rho_{1}
\end{array}
\right. \\
\makebox[1cm]{c)} & \makebox[10cm]{
The result are the same as in b). The quantum states } \\
& \makebox[10cm]{b), c) are identical from the point of view of the detector.}
\end{array}
$$

\subsection{Properties of the detector D II }
\label{detector_d2}

The system working as a detector D II should satisfy the following 
condition: to be able to distinguish at least two different quantum states, 
the efficiency of the detector should be as big as possible.
Taking under consideration the above condition I shall use the system 
described in the section [\ref{twodetector}] which satisfy:
\begin{itemize}
\item $ k_{1} = n_{1}= n $ should be big enough to satisfy the conditions of
the section [\ref{detector1}]
\item $ k_{2} = n_{2} = 0 $
\end{itemize}

\subsection{Implementation of the system.}
\label{examplesys}

In the following I consider the properties of the transmition thorough such 
a system. \newline
I make the assumption:
\begin{itemize}
\item The source generates the signal $ \rho$ such that
$ Tr(e_{1} \rho ) \geq 0,8 $ (or $ Tr( e_{2} \rho ) \geq 0,8 $) (the
choice depends on the value which is being transmitted).\footnote{ This 
assumption say that the source generate a good quality signal but the noise 
is allowed also.}
\item The parameters of the detectors should be chosen in such a way that 
the real efficiency of the detectors achives $ 90 \% $ of the maximum one.
\item The classical subsystem satisfy the initial condition [\ref{pocz}].
\end{itemize}
Under the above condition I receive the probabilities of the registration of 
the signal:

\begin{eqnarray}
D I & & p_{1} (t) = \frac{1}{2} ( 1 - e^{-2kt} ) \rho_{1} \\
\label{rown29a}
D II & & 
\begin{array}{l}
p_{1} (t) = ( 1 - e^{- n^{2} t} ) \rho_{1} \\
p_{2} (t) = ( 1 - e^{- n^{2} t} ) \rho_{2}
\end{array}
\label{rown29b}
\end{eqnarray}

I would like to stress that this equation describe the evolution of the 
probabilities of the registration of the state. What more the probability of 
an event may be specified after an infinity serious of experiments.
However the probability of the registration of a state may be specified with 
some level of confidence. I define the confidence level as a $90 \%$, this 
means that with the probability of $90 \% $ the measured value is the real 
value.

The aim of the work of the detectors:
\begin{description}
\item{D I} This detector check if the signal has been
generated.\footnote{ I understand as a signal a quantum state with some
(specified) properties.} So I have to designate how many states should
be generated by the source in order to be sure that the detector
register at least one state.
\item{D II} It recognize the kind of the signal. In order to do this
job, the detector should designate if the value $\acute{\rho_{1}}$
belongs to the interval $( \rho_{1} - a, \rho_{1} +a)$ within the
confidence level.
\end{description}

\subsubsection{Detector D I.}

For producing an effect by the system D I is required a very small number
of states. I assume that the act of generation of a state is an
independent event, so the probability of registration at least one of the
generated states could be found as a probability of complementary event
to the no registration of the states. I have:
\be
P( \mbox{no registration} ) + P( \mbox{registration} ) =1
\label{rown29}
\ee
and from [\ref{rown29a}]
\be
P( \mbox{no registration} ) = ( \frac{1}{2} ( 1 - e^{-2kt_{0} } ) \rho_{1}
)^{n}
\label{rown30}
\ee
where: \newline
$ n$ -- number of generated states, \newline
$ t_{0} $ -- time of evolution of coupled systems. \newline
Applying equations [\ref{rown29}] [\ref{rown30}] we can find out, that it
is enough, with the assumed confidence level, to generate 4 state to be sure
that the detector \nel
D I confirm the fact of sending the signal.

\subsubsection{Detector D II.}

The situation of the detector D II is more complicated than with D I.
It is not enough to confirm the fact of receiving the state, but the
value of $\rho_{1}$ should be estimated, otherwise we do not know what
kind of signal is being received.

I evaluate the probability of registration the state, which has the
coefficient of the vector $e_{1}$ belonging to the interval between
$ w_{1} = \rho_{1} - a $, \newline $ w_{2} = \rho_{1} +a$. \newline
$a$ -- the measuring accuracy, \newline
$\rho_{1}$ -- the real value, \newline
$ w_{1}, \: w_{2}$ -- respectively the infimum and supremum of the
interval.

The following value is obtained trom the experiment:
\be
\acute{p_{1}} (t) = \frac{i}{m}
\ee
$i$ -- number of state registered by the detector D II as a particular
logical value (let it be "1"), \newline
$m$ -- number of the states generated by the sender in order to send the
message.\footnote{It is very important to know the number of states
generated by the sender in time.} \newline
I introduce two auxiliary quantity:
\begin{description}
\item{$p_{1}^{-} (t_{0}) $ --} the probability of registration of the state
during the measurement in time $t_{0}$, where the decomposition of
$ \rho $ is $\rho = w_{1} e_{1} + \sum_{i=2} \rho_{i} e_{i} $,
\item{$p_{1}^{+} (t_{0}) $ --} the probability of registration the state
$\rho$, where
\mbox{$ \rho = w_{2} e_{1} + \sum_{i=2} \rho_{i} e_{i} $}.
\end{description}
The expectation value is respectively:
\be
\begin{array}{l}
i^{-} = m \: p_{1}^{-} (t_{0}) =
m \: p_{1} (t_{0}) - (1 - e^{-n^{2} t_{0}} ) a\: m \\
i^{+} = m \: p_{1}^{+} (t_{0}) =
m \: p_{1} (t_{0}) + (1 - e^{-n^{2} t_{0}} ) a \: m
\end{array}
\label{rown31}
\ee
I solve the following two aspects:
\ben
\item What is the smallest number of states required for obtaining such
$ \acute{p_{1}}$ that estimated $ \rho_{1} \in ( w_{1} , w_{2} )$ ?
\item What is the smallest number of states, that with the certain
confidence level is possible to say that  $ \rho_{1} \in ( w_{1} , w_{2} )$ ?
\een
{\bf The point a) } \nel
It is possible to obtain the result which belong to the specified
interval, when the expectation value $ i^{-}, \: i^{+}$ satisfy the
condition:
$$
i^{-} \leq [ i^{+}]
$$
$ [ i^{+}] $ -- the integer part of $ i^{+} $ \nel
so the intersection $ [ i^{-}, \, i^{+} ] \cup {\cal N} \neq 0 \!\!\! / $
\nel
$ {\cal N} $ -- the set of integer numbers. \nel
The above condition I rewritte in the more usefull form:
$$
i^{+} - i^{-} \geq 1
$$
so for the detector D II I obtain:
\be
m \geq \frac{1}{2(1- e^{-n^{2} t_{0} } )a }
\label{rown32}
\ee
The value obtained in [\ref{rown32}] specify the lower limit of the
number of generateg states required to change the classical state of the
detector. It can be seen that after a few generated states it is
possible to receive the message properly. \nel
{\bf The point b).} \nel
I use the Bernoulli distribution for evaluating the probability of
registration of "$ i $" states among "$ m $" received is:
\be
P ( i, m ) = \left( \begin{array}{c} m \\ i \end{array} \right)
( p_{1} (t) )^{i} ( 1- p_{1} (t) )^{m-i}
\ee
For obtaining the probability of evaluation $ \acute{p_{1}} \in (
p^{-}_{1} , p^{+}_{1} ) $ The distribution should be sum up over the
advantageous events. \nel
I introduce the function: \nel
$[x]_{-}$ -- the integer number which $ [x]_{-} \leq x $ \nel
$[x]_{+}$ -- the integer number which $ [x]_{+} \geq x $ \nel
so the probability is:
\be
P(m) = \sum_{i=[i^{-} ]_{+}}^{[i^{+}]_{-}} P(i, \, m)
\label{rown33}
\ee
{\bf Remark} \nel
It is noticable, that with the increasing number of the generated states
the probability of obtaining the proper value does not have to increase.
It may happen that dispite increasing the number of sended states the
probability decrease. This fact should be taken under consideration
during the real experiment.

\subsubsection{ Examples:}
\label{examples}

In the follwing examples I assume: \nel
$ \rho_{1} = 0,8 $, \nel
$ t_{0} $ is such that $ 1- e^{ -n^{2} t_{0}} = 0.9 $ \nel
the measuring accuracy of $ \rho_{1}$: $a=0,05$ \nel
so I obtain: \nel
from the equation [\ref{rown32}] the smallest number of states, which
should be generated:
$$
m \geq 12
$$
In this case the expectation value $ i^{-}, \: i^{+} $ are :
$$
\begin{array}{l}
i^{-} = 8,1 \\
i^{+} = 9,18
\end{array}
$$
We see that the only advantageous event is to registrate 9 states, then
\nel $ \acute{p_{1}} = 0,75$ and $ p_{1} = 0,72$ \nel
The probability of obtaining this result is:
\be
P(12) = P(9,12) =
\left( \begin{array}{c} 12 \\ 9 \end{array} \right) ( 0,72 )^{9}
(0,28)^{3} = 0,25
\ee

Example to the remark. \nel
For $m=12$ the adventegous event is to measure 9 states and the
probability of such event is: \nel
$ P(12)=0,25$ \nel
For $m=15$ the adventegous event is to measure 11 states, then : \nel
$P(15)=0,22$ \nel
We see that the probability of obtaining the proper result decrease
dispite increase the number of the generated states.

Achiving the confidence level higher then 60\% is possible for $m=62$.
The advantegous events are: \nel
$ i= \{ 42, 43, 44, 45, 46, 47 \}$ and $ P(62) = 0,603$ \nel
It is interesting that the efficiency of the detector is not extremally
important because for the detector working with the efficiency 0,45 (it
is half of the efficiency assumed above), for $m=66$ the system achives
similar confidence level.
In this case: \nel
$ i= \{ 21, 22, 23, 24, 25, 26 \}$ and $P(66) = 0,56$ \nel
So very close to the above example.

\subsection{Conclusion}
\label{end}

This calculation could be simplyfied by replacing binominal distribution
by normal distribution, but this can be done only for big number of
events, whereas the aim of the work was to estimate the smallest number
of states required for achiving the assumed quality of transmition.

The investigation presented in whole work does not exhaust the range of
improvement of such a transmition e.g. the control signal can be
introduced in order to increase the confidence level or resistance to
evasdropping; use the sophisticated way of coding including several
possibilities of quantum error corection codes
\cite{2,shor,shora,zurek} and so on. The number of possible
improvement is great.

The last question which I'd like to consider is the resistance to noise.
In the previous example I've used the signal which had as its part a
20\% of noise and dispite of it the transmition was possible. I admite
that usage of more then 2 state coding may considerably increase the
speed of the transmition. The boundaries of n--state codding may be the
properties of the quantum system (the Hamiltonian of this system) and of
the channel.
I've omitted the role of the channels, but in the farther investigation
its role should be considerd also.

\section{Aknowledgments}

I'd like to thenk prof. A. Jadczyk for his helpful discussions and hints
during the work on this subject.

\section{Appendix}
\label{appendix}
\licznik

The maximum  efficeincy of the n--state detector.

I assume the following coupling operator $ V_{i}$:
\be
V_{i} = \sqrt{k} \left( \begin{array}{ccccc}
0 & \cdots & e_{i} & \cdots & 0 \\
0 & \cdots & 0 & \cdots & 0 \\
\vdots &  & \vdots &  & \vdots \\
0 & \cdots & 0 & \cdots & 0
\end{array}
\right)
\label{a1}
\ee
Where $e_{i}$ are projectors operators. \nel
The operator [\ref{a1}] satisfy the conditions [\ref{war1}] and
[\ref{war2}]. The equation [\ref{rown1}] in this case take the form:
\be
\left\{ \begin{array}{l}
\dot{\rho}_{0} (t) = -i[H, \rho_{0} (t) ] -
\frac{1}{2} k ( \sum e_{i} e_{i}
\rho_{0} (t) + \rho_{0} (t) \sum e_{i} e_{i}) \\
\dot{ \rho}_{i} (t) = -i[H, \rho_{i} (t) ] + k e_{i} \rho_{0} (t) e_{i}
\end{array}
\right.
\label{a2}
\ee
Using the trace operations:
\be
\left\{ \begin{array}{l}
\dot{p}_{0} (t) = -k \, Tr \, \sum e_{i} \rho_{0} (t) \\
\dot{p}_{i} (t) = k \, Tr \, e_{i} \rho_{0} (t)
\end{array}
\label{a3}
\right.
\ee
I assume that the commutator of the Hamiltonian and $\rho$ does not
change the decomposition of $\rho$. So the equations [\ref{a1}] may be
rewritten in the form similar to [\ref{rown9b}].
\nel
Let
\be
\rho_{0} (0) = a_{j} e_{j}
\ee
So I obtain
\be
\left\{ \begin{array}{l}
\dot{p}_{0} (t) = -k \, Tr \, \sum a_{j} (t) e_{j}  \\
\dot{p}_{j} (t) = k \, Tr \, a_{j} (t) e_{j} \\
\dot{p}_{i \neq j} = 0
\end{array}
\label{a4}
\right.
\ee
The solution of the equations [\ref{a4}] are:
\be
\left\{ \begin{array}{l}
p_{0} = A e^{-kt} \\
p_{j} = C - B e^{-kt} \\
p_{i \neq j } = const
\end{array}
\right.
\label{a5}
\ee
Applying the initial condition [\ref{pocz}]: \nel
$ A = 1 $ \nel
Assuming that $ p_{j} ( \infty ) = 1$ (the ideal efficiency): \nel
$ C = 1 $ \nel
and from the prbability theory: \nel
$B = A = 1$ \nel
So the solution take the form:
\be
\left\{ \begin{array}{l}
p_{0} = e^{-kt} \\
p_{j} = 1 - e^{-kt} \\
p_{i \neq j } = 0
\end{array}
\right.
\label{a6}
\ee
The asymptotic value of the efficiency of the detector is one, of course
when the initial state has the required form.

\end{document}